\documentclass[12pt,psfig]{article}
\usepackage{amssymb}
\usepackage{amsmath}
\usepackage{tensor}
\usepackage{multirow}
\usepackage[a4paper]{geometry}
\usepackage{cite}

\usepackage{color}

\newcommand{\sm}{\textrm{-}}

\def\be{\begin{equation}}
\def\ee{\end{equation}}
\def\bea{\begin{eqnarray}}
\def\eea{\end{eqnarray}}
\def\f#1#2{\frac{#1}{#2}}
\begin{document}
\begin{titlepage}
\title{{\bf Galileons from Lovelock actions}} \vskip2in

\author{
{\bf Karel Van Acoleyen and Jos Van Doorsselaere\footnote{\baselineskip=16pt E-mail: {\tt
karel.vanacoleyen@ugent.be, jos.vandoorsseleare@ugent.be
}} }
\hspace{3cm}\\
$$ {\small Department of Physics and Astronomy, Ghent
University,}\\
{\small Krijgslaan 281, S9, 9000 Gent, Belgium}.
}

\date{}
\maketitle
\def\baselinestretch{1.15}

\begin{abstract}
\noindent
We demonstrate how, for an arbitrary number of dimensions, the Galileon actions and their covariant generalizations can be obtained through a standard Kaluza-Klein compactification of higher-dimensional Lovelock gravity. In this setup, the dilaton takes on the role of the Galileon. In addition, such compactifications uncover other more general Galilean actions, producing purely second-order equations in the weak-field limit, now both for the Galileon and the metric perturbations.

\end{abstract}

\thispagestyle{empty} \vspace{5cm}  \leftline{}

\vskip-19cm  \vskip3in

\end{titlepage}
\setcounter{footnote}{0} \setcounter{page}{1}
\newpage
\baselineskip=20pt

\section{Introduction}
Inspired by the decoupling limit of the DGP model, Galileon models were introduced as scalar field models whose equations on flat space are purely second-order \cite{NRT} (but see also\cite{Fairlie}). This demand translates to a symmetry - the so called Galilean symmetry - which is quite restrictive: apart from the linear term and the standard quadratic kinetic term, there are only three more Galileon terms that are allowed in the action (for space-time dimension $D=4$). Together with the flat-space second-order equations, this makes these models nice from a phenomenological standpoint, both at the classical and quantum level \cite{NRT,EFT}. And so far Galileon(-like) models have been studied in the context of early \cite{Inflation} and late \cite{Latetime,VanAcoleyen:2009kb} time cosmology and as a possible origin for a fifth force \cite{fifthforce}. Due to the Galilean symmetry, such a fifth force would still largely respect the equivalence principle \cite{equivalence}.

Of course, to consider for instance the cosmology of these models, one needs to couple the Galileon $\pi$ to the gravity field $g_{\mu\nu}$. As was shown in \cite{DEV}, in general the minimally coupled terms lead to third-order equations. This is undesirable, as higher derivatives typically lead to instabilities. However, it was also shown how the higher derivatives can be eliminated through the inclusion of several precisely tuned non-minimal curvature coupling terms. Subsequently this construction was put in a remarkably elegant form and generalized to arbitrary dimension in \cite{DDE}, by Deffayet, Deser and Esposito-Farese (DDE). As the authors noticed, the general form of the Galileon actions, was 'tantalizing' reminiscent of that of Lovelock gravity. And indeed a connection was found. In \cite{dRT} it was shown how every 4D fully covariant Galileon term arises in the appropriate limit of the action of a probe brane, by considering all nonzero Lovelock invariants, both on the brane and in the 5D bulk. The Galileon is in this context identified with the position modulus of the brane.

In this paper we provide a different connection with Lovelock gravity, now for arbitrary dimension $D$. To see how the connection could emerge, let us recall that the Lovelock invariants are the unique scalars, constructed solely out of the metric, that generate Lagrangian equations which are {\em at most} of second order \cite{Lovelock:1971yv}. And for a given Lovelock invariant, at the lowest non-trivial order in the weak-field expansion in the metric perturbations $h_{\mu\nu}$ $(=g_{\mu\nu}\!-\!\eta_{\mu\nu})$, the equations are purely second-order. This is also precisely the case for the fully covariant Galileon actions, for which the equations are at most of second order, with purely second-order equations at lowest order (in this case zero) in the weak-field expansion. It is then evident to look for a connection with Lovelock gravity, by considering a $D+N$ higher-dimensional Lovelock theory, and singling out some $D$-dimensional scalar $\pi$ in the extra-dimensional components of the metric. The resulting equations that arise after compactification of the extra dimensions, will then be at most second-order, both in $\pi$ and in the metric $g_{\mu\nu}$ of the uncompactified dimensions. Furthermore one would expect some of the terms in the equations to be {\em Galilean}, in the sense that they are purely second-order in the weak-field limit.\footnote{Such purely second order terms necessarily have the Galilean symmetry for all fields involved. Which in this case means: $\pi\rightarrow \pi+c+c_\mu x^\mu$ and $h_{\alpha\beta}\rightarrow h_{\alpha\beta}+d_{\alpha\beta}+d_{\alpha\beta\mu}x^\mu $.} This is indeed what we find. In section \ref{sec3}, we write down the decomposition of the action for the standard Kaluza-Klein ansatz for the metric, with $\pi$ identified as the dilaton that controls the size of the extra dimensions. We find that different terms in the action independently lead to second-order equations for $\pi$ and the $D$-metric $g_{\mu\nu}$. In section \ref{sec4} we discuss the different terms at the level of the equations which we then classify in section \ref{sec5}. Some of the obtained terms are indeed Galilean, and amongst these we find precisely the fully covariant Galileon actions of \cite{DDE}, in two equivalent versions. But we also uncover other Galilean actions, that in the weak-field limit are of nonzero order in $h$, and that can be thought of as the interpolations between the original scalar field Galileon actions and the pure gravity Lovelock actions. In section \ref{D=4} we specialize to the case $D=4$ and write down explicitly all possible Galilean terms that can be generated by Lovelock compactifications. But before delving into compactifications of Lovelock-gravity, let us first review the results of \cite{DDE}.

\section{Generalized Covariant Galileons}\label{sec2}
As we already mentioned, DDE were able to condense their findings in a remarkably elegant formalism. For arbitrary dimensions $D$ they found the following form for the fully covariant Galilean actions in arbitrary dimensions:
\be\mathcal{L}^n=\sum^{\frac{n-1}{2}}_{p=0}\mathcal{C}^n_p\mathcal{L}^n_p\quad\mathrm{with}\quad \mathcal{L}^n_p=\mathcal{A}_{(2n)}\pi_1\pi_2\mathcal{S}(q)\mathcal{R}(p)\,,\ee
for certain coefficients $\mathcal{C}_p^n$.
We adopt their notation throughout the paper, except for the assignment of the upper-index in $\mathcal{L}$, where DDE used the degree in $\pi$ (in this case simply $n+1$) rather than $n$ to label the different actions. We have further
\be \mathcal{A}_{(2n)}=\delta^{\mu_1\mu_3\cdots\mu_{2n-1}}_{[\nu_2\nu_4\cdots\nu_{2n}]}g^{\nu_2\mu_2}g^{\nu_4\mu_4}\cdots g^{\nu_{2n}\mu_{2n}}\,,\label{A2n}\ee
such that all other, contracted, indices are lower ones. We drop the 'covariant' semicolons and identify $\mu_i \sim i$:
\be \pi_i=\pi_{;\mu_i}\,\,,\quad \pi_{ij}=g_{\mu_j\nu_j}\pi_{;\mu_i}^{\ \nu_ j}\,\,,\ldots\,.\ee
Also we use the functions
\begin{eqnarray}
\mathcal{S}(q)&=&\prod_{i=a}^{q+a-1}\pi_{\mu_{2i-1}\mu_{2i}}\,,\\
\mathcal{R}(p)&=&(\pi_{\nu}\pi^{\ \nu})^{p}\prod_{k=b}^{p+b-1}R_{\mu_{4k-3}\mu_{4k-1}\mu_{4k-2}\mu_{4k}}
\,,\end{eqnarray}
for which the starting point of the index-counting is always appropriately chosen. Since $n$ is the total number of anti-symmetrized (odd or even) indices, we have $n=1+q+2p$.

As DDE notice, the appearance of $\mathcal{A}_{(2n)}$ seems to indicate a very close relation to Lovelock-gravity. It is this relation that we will make explicit in this paper. 

\section{Decomposing a Lovelock action}\label{sec3}  As we discussed in the introduction, to find a link between Galileon models and Lovelock gravity, it seems suitable to study Lovelock gravity in higher dimensions. Therefore we will study the equations that are generated by a certain Lovelock invariant in a $D+N$-dimensional space-time. For the metric we take a standard Kaluza-Klein anzats that factorizes the dimensions:
\begin{equation}
g_{AB}=\left(\begin{array}{cc} g_{\alpha\beta}(x^{\mu})&0\\0&e^{\pi(x^\mu)}\gamma_{ab}(x^i)\end{array}\right)\label{metric}
\,,\end{equation} with $g_{\alpha\beta}$ and $\pi$ depending on the first $D$ coordinates, while $\gamma_{ab}$ depends on the $N$ extra-dimensional coordinates. The use of greek and latin indices will consistently indicate the separation between the two, while capital indices will refer to both.  We take an arbitrary background-metric $\gamma_{ab}$ for the auxiliary space and we will consider the equations for $\pi$ and $g_{\alpha\beta}$ that are generated by the $d$-th order Lovelock invariant:\begin{equation}
L^{N+D}_{(d)}=\delta_{[B_1\ldots B_{2d}]}^{A_1\ldots A_{2d}}R_{A_1A_2}^{\quad\,\,\,  B_1B_2}\cdots R_{A_{2d-1}A_{2d}}^{\quad\,\,\,\quad\,\,\, B_{2d-1}B_{2d}}\,.\label{LL}
\end{equation}
To this end we need its decomposition in the lower-dimensional components for the metric (\ref{metric})\,.  With some work, that we relegate entirely to appendix \ref{decomp}, we find:
\begin{eqnarray}
L_{(d)}^{N+D}&=&\sum_r e^{-r\pi}L^N_{(r)}\sum_n\mathcal{\tilde C}^n\sum_p\left(\mathcal{D}^n_p\mathcal{K}^{n}_{p}+\mathcal{C}^n_p\mathcal{L}^{n}_{p}\right)\,,\nonumber\\&\equiv&\sum_r e^{-r\pi}L^N_{(r)}\sum_n\mathcal{\tilde C}^n(\mathcal{K}^{n}+\mathcal{L}^{n})\,,\label{ff}\eea
with $L^N_{(r)}$ the $r$-th order Lovelock from $\gamma_{ij}$ (with the convention that $L^N_{(0)}=1$) and
\begin{eqnarray}
\mathcal{K}^{n}_{p}&=&\mathcal{A}_{(2n)}\mathcal{S}(q+1)(\pi_{\nu}\pi^{\ \nu})^{d-n-r}\mathcal{R}(p)\nonumber\,,\\
\mathcal{L}^{n}_{p}&=&\mathcal{A}_{(2n)}\pi_{1}\pi_{2} \mathcal{S}(q)(\pi_{\nu}\pi^{\ \nu})^{d-n-r}\mathcal{R}(p)\,,\nonumber\\
\mathcal{C}^n_{p}=\frac{n-2p}{2}\mathcal{D}^n_p&=& \frac{(-2)^{-3p}}{p!(d-n+p-r)!(n-2p-1)!}\,,\nonumber\\
\mathcal{\tilde{C}}^n&=&\frac{(-2)^{2n+r-d}}{2}\frac{d!}{r!}\frac{(N-2r)!}{(N+n-2d)!}\,.\label{decompterms}
\end{eqnarray}
Notice that the power of $\pi$ for all terms in $\mathcal{K}^n$ and $\mathcal{L}^n$ is fixed, with $2(d-r)-n$ powers of $\pi$ for the former and $2(d-r)-n+1$ for the latter. Notice also that the number of extra dimensions $N$ only enters the overall factor $\tilde{\mathcal{C}}^n$ and that the Lovelock orders $d$ and $r$ only enter $\mathcal{K}^n$ and $\mathcal{L}^n$ in the combination $(d-r)$.

We still need to specify the summation bounds in (\ref{ff}). The $r$-summation runs from 0 to $d$. The $n$ and $p$-summations run simply over all possible terms that are nonzero. This gives different bounds for the $\mathcal{K}$ and $\mathcal{L}$ terms, respectively: \bea 0\leq n\leq \min{(2(d-r),D)}\quad\mbox{,}\quad \max({0,n-(d-r)})\leq p\leq{ n/2}\,;\nonumber\\
1\leq n\leq \min{(2(d-r)-1,D)}\quad\mbox{,}\quad \max({0,n-(d-r)})\leq p\leq{ (n-1)/2}\,.\label{bounds}\eea

By the Lovelock construction, the full equations that arise from (\ref{LL}) are (at most) second-order. But for our purposes it is interesting that the equations that arise from each $\mathcal{K}^n$ and $\mathcal{L}^n$ term in (\ref{ff}) separately, are actually by themselves also second-order. To demonstrate this, one could simply write down the explicit equations that derive from these terms. We will take a shorter path, that exploits the freedom for the extra-dimensional background-metric $\gamma_{ab}$ and the number of extra dimensions $N$.

Let us first notice that upon a rescaling of the background-metric $\gamma_{ab}$, the different coefficients $L^N_{(r)}$ in (\ref{ff}) scale differently. This implies that there can not be any cancellation of higher derivatives between different terms in the $r$-summation, or, in other words, that these terms independently lead to second-order equations.

Secondly, we prove that the second-order nature of the action does not change by adding (or canceling) any pre-factor $f(\pi)$. Indeed, for a Lagrangian of the form $f(\pi)\times(\ref{ff})$, the only possible extra higher order terms could come from a second order derivative - that arises through variation with respect to $\pi$ in $\mathcal{S}(q)$ or with respect to $g_{\mu\nu}$ in $\mathcal{R}(p)$ - with one derivative working on the pre-factor, and the other working on the remaining factors in $\mathcal{S}(q)$ or $\mathcal{R}(p)$. Because of the anti-symmetry in $\mathcal{A}_{(2n)}$ and the Bianchi-identity for the Riemann-tensor, the latter terms are indeed only second-order.

So we can drop the exponential pre-factor to analyze the second-order nature of the different terms in (\ref{ff}). Without this pre-factor, there is a one to one correspondence between terms with a fixed power of $\pi$ in the action and in the equations. And it is clear that the different powers of $\pi$ in the equations are all second-order - higher derivatives can not cancel out between terms with different powers. In the action, the terms with a certain power of $\pi$ have the form $\mathcal{\hat C}^n(N)\mathcal{K}^n+\mathcal{\hat C}^{n+1}(N)\mathcal{L}^{n+1}$.  As the two coefficients vary differently with $N$, both terms indeed lead independently to second-order equations. This concludes the argument.

\section{The different terms in the equations}\label{sec4}
A final subtlety in distinguishing the different independent types of terms at the level of the equations, is that the different $\mathcal{K}^n$ and $\mathcal{L}^{n+1}$ terms that we discussed above, are actually identical up to a total derivative. As we show in appendix \ref{divergence}: \begin{equation}
\frac{2(d-r)-n}{2}\mathcal{K}^n=2\mathcal{L}^{n+1}+
\nabla_1\left(\pi_2\mathcal{A}_{(2n)}\sum_p\mathcal{C}_p^n\,\mathcal{S}(q)\mathcal{R}(p)(\pi_\nu\pi^\nu)^{d-n-r}\right)\,,\label{KL}
\end{equation} with the $p$-summation again over all possible nonzero terms: \be \max({0,n-(d-r)})\leq p\leq{ (n-1)/2}\,.\ee We can use this relation to substitute the $\mathcal{K}$ terms with $\mathcal{L}$ terms in the full action, except for the case $n=2(d-r)$, with the $\mathcal{K}$ term the order $d-r$ Lovelock invariant, $\mathcal{K}^{2(d-r)}=L^D_{(d-r)}$, as can be seen from (\ref{decompterms}). Up to surface terms, the full action that arises from the compactification of the $d$-th Lovelock invariant then finally becomes:
\begin{equation}
S=\sqrt{-g}e^{\frac{N}{2}\pi}L^{N+D}_{(d)}=\sqrt{-g}\sum_r L^N_{(r)}e^{(\frac{N}{2}-r)\pi}\left(\tilde{\mathcal{C}}^{2(d-r)}L^D_{(d-r)}+
\sum_n\mathcal{\hat C}^n\mathcal{L}^n\right)\,,\label{action}\end{equation}
with now: \be \mathcal{\hat C}^n=\frac{2d-n-N}{\left(2(d-r)-n\right)\left(2(d-r)-n+1\right)}\tilde{\mathcal{C}}^n\,. \ee

\section{Classification of the different terms}\label{sec5}
Equation (\ref{action}) shows all the different terms in the action, that arise in a standard Kaluza-Klein compactification (\ref{metric}) of the order $d$ Lovelock invariant. The Lovelock-invariant $L^D_{(d-r)}$ and the terms $\mathcal{L}^n$ each give rise to equations that are independently second-order, and as we argued above, this holds regardless of the pre-factor. Furthermore, it is clear now that the different terms do give different equations as they all have a different power of $\pi$.

Rather than looking at a particular compactification of a particular Lovelock invariant, we will now classify all possible different terms (modulo the pre-factor) for a certain dimension $D$, that can arise in arbitrary Lovelock compactifications. Recall that both $L^D_{d-r}$ and $\mathcal{L}^n$ in (\ref{action}) are independent of the number of extra dimensions $N$ and that they only depend on $n$ and $d-r$. In further expressions we will take $r=0$, no expressions will be omitted in this way. Now, first of all we have of course all the non-zero Lovelock invariants $L^D_{(d)}$ that can appear. With the condition $2d\leq D$, this gives $D/2$ or $(D-1)/2$ different terms, for an even or odd number of dimensions $D$ respectively.\footnote{Omitting the cosmological constant $L^D_{(0)}$.}

Secondly, we find the terms $\mathcal{L}^n$ that all involve two or more powers of the scalar $\pi$.  We base the classification of these terms on their Galilean character. As we explained in the introduction, we qualify a term as Galilean, if it leads to purely second-order equations in the weak-field limit. We therefore count the number of fields and derivatives in the weak-field limit, for every term $\mathcal{L}^n_p$ in $\mathcal{L}^n=\sum_p\mathcal{C}^n_p\mathcal{L}^{n}_{p}$. The total number of fields for Galilean terms will be one more than half the number of derivatives, $n_f=n_d/2+1$, as this gives two derivatives per field in the equations. Putting $R_{\mu_1\mu_2\mu_3\mu_4}\sim \partial^2 h$, we find from (\ref{decompterms}) for every $\mathcal{L}^n_p$ term:  $2d-n+1$ factors of $\pi$, $p$ factors of metric-perturbations $h$, and a total number of $2d$ derivatives. This means that only the terms $\mathcal{L}^n_p$ with $p=n-d$ lead to purely second-order equations in the weak-field limit. Inspection of the summation bound  (\ref{bounds}) for $p$ leads us then to the conclusion that the $\mathcal{L}^n$ terms with $1\leq n<d$ are non-Galilean, with first-order derivatives in the equations already in the weak-field limit\footnote{These terms still have the ordinary shift-symmetry $\pi\rightarrow
\pi+c$.}. From the bound (\ref{bounds}) on $n$, one finds that there exist an infinite number of different terms of this type.

The more interesting Galilean terms are those with $n\geq d$. Indeed, for those terms, the $p$-summation starts at $p=n-d$, and as we just argued, $\mathcal{L}^n_{n-d}$ does give rise to purely second-order equations in the weak-field limit. Notice that for a given $n$, the other terms $\mathcal{L}^n_{p}$ in the $p$-summation come with more powers of curvature, and are therefore of higher order in the weak-field expansion.

We can further differentiate the Galilean terms that exist on flat space from those that do not. Since the degree in $h$ of $\mathcal{L}^n_{n-d}$ is $n-d$, the only terms belonging to the first category are the $\mathcal{L}^n$'s with $n=d$. From (\ref{bounds}) we find the bound $1\leq n\leq D$ for such terms, leading to a total of $D$ different Galilean terms in $D$ dimensions, that are not-trivial on flat space. These are the terms that were studied by DDE in \cite{DDE}, and we indeed recover the very same expressions. On flat space we have $\mathcal{L}^d\sim \mathcal{L}^d_0$, which is precisely the form of the general flat-space Galileon that was put forward by DDE. The other terms $\mathcal{L}^d_p$ in $\mathcal{L}^d$ are trivial on flat space, but as DDE found, they are required to maintain second-order equations for both the metric and $\pi$ equations on a general background. In our setup these terms follow automatically from the Lovelock construction, with, up to an overall factor $(d-1)! $, the same coefficients $\mathcal{C}^d_p$ that were found by DDE. Furthermore, the Lovelock construction uncovers an equally elegant, equivalent form for the fully covariant Galileons since by (\ref{KL}) we have that $\mathcal{L}^{d}\cong\mathcal{K}^{d-1}$ at the level of the equations. The starting flat-space term in the summation $\mathcal{K}^{d-1}=\sum_p\mathcal{D}^{d-1}_p\mathcal{K}^{d-1}_{p}$ now reads: \be \mathcal{K}^{d-1}_{0}=\mathcal{A}_{(2d-2)}\mathcal{S}(d-1)(\pi_{\nu}\pi^{\nu})\,, \ee and one could actually construct the other terms $\mathcal{K}^{d-1}_p$ (with the proper coefficients) in a similar procedure as followed by DDE, by requiring the cancellation of all higher order derivatives.

The Lovelock setup also reveals flat-space trivial Galilean terms (for $n>d$). These have $n-d>0$ powers of curvature at the lowest order in the weak-field expansion $\mathcal{L}^n\sim \mathcal{L}^n_{n-d}$. This is a new class of scalar-tensor couplings, and we can think of them as interpolations between the original scalar field Galilean actions and the pure spin 2 Lovelock actions. From the bound (\ref{bounds}) on $n$, one finds that there exist $D(D-2)/4$ or $(D-1)^2/4$ different terms of this type, for $D$ even or odd.

\section{Galilean terms for D=4}\label{D=4}
We will now specialize to the case $D=4$, and explicitly write down all Galilean terms that can appear in Lovelock compactifications. Let us start with the original scalar field covariant Galileons of \cite{DEV,DDE}. As we discussed above, we find them appearing in two equivalent forms: the $\mathcal{L}$-form that was found by DDE and a new $\mathcal{K}$-form. The explicit expressions are simpler in this new form, since they involve less anti-symmetrized indices. From eqs. (\ref{ff},\ref{decompterms}) we find (recall that we take $r=0$):
\bea
\mathcal{K}^0 (d=1) &=& 2(\partial \pi)^2\,,\label{K0}\\
\mathcal{K}^1 (d=2) &=& 2\Box\pi (\partial\pi)^2\,,\label{K1}\\
\mathcal{K}^2 (d=3) &=& (\partial \pi)^2\left((\Box\pi)^2-\pi_{\mu\nu}\pi^{\mu\nu}\right)-\f{1}{4}(\partial\pi)^4 R\,,\\
\mathcal{K}^3 (d=4) &=& \f{1}{3}(\partial\pi)^2\left((\Box\pi)^3-3\pi_{\mu\nu}\pi^{\mu\nu}\Box\pi +2\pi_{\mu}^{\,\,\nu}\pi_\nu^{\,\,\rho}\pi_\rho^{\,\,\mu}\right)
+\f{1}{2}(\partial\pi)^4\pi_{\mu\nu}G^{\mu\nu}\,.\label{K3}\eea
Upon one partial integration of the last term in (\ref{K3}), these are exactly the original $D=4$ covariant Galileons that were obtained in \cite{DEV}.

The Lovelock-terms form another group of Galilean terms. For $D=4$, the two nonzero terms are the Ricciscalar and the Gauss-Bonnet invariant, that appear multiplied by some pre-factor $f(\pi)$. Focussing on Galilean terms, we can write three independent terms that come from the Lovelock's:
\bea
L^4_{(1)}&=&R\,,\label{R}\\
\pi L^4_{(1)}&=&\pi R\,,\label{piR}\\
\pi L^4_{(2)}&=&\pi \left(R^2-4R_{\mu\nu}R^{\mu\nu}+R_{\mu\nu\rho\sigma}R^{\mu\nu\rho\sigma}\right)\,.\label{piGB}
\eea
Indeed, one can easily verify that any Lovelock-term leads to purely second-order equations in the weak-field limit, both by itself and multiplied by a single power of $\pi$.\footnote{Note that all Lovelock-invariants are total derivatives at leading order in the weak-field expansion. It is only for the next-to-leading-order term that the equations are non-trivial. Counting the number of fields and derivatives as in the previous section, we indeed have $n_f=n_d/2+1$ for this term. Likewise, for the Lovelock invariant multiplied by $\pi$ we have the same relation, now at leading order in the weak-field expansion. } For the Gauss-Bonnet invariant only the latter case is non-trivial since the invariant by itself is a total derivative. The well known scalar-tensor couplings (\ref{piR}) and (\ref{piGB}), appear for instance in Brans-Dicke gravity (in the Jordan-frame), $f(R)$ gravity and Gauss-Bonnet gravity. The Galilean character of the Gauss-Bonnet coupling (\ref{piGB}) was already exposed in \cite{VanAcoleyen:2009kb}, in the context of the cosmological backreaction for Gauss-Bonnet gravity.

Now, it is often stated that (\ref{piR}) and (\ref{piGB}) form the only consistent non-minimal scalar-tensor couplings (modulo redefinitions $\pi\rightarrow f(\pi)$), in the sense that they do not generate higher derivative equations. But, as we discussed in the previous section, the Lovelock construction uncovers another class of scalar-tensor couplings, that lead to second-order equations, with purely second-order equations in the weak-field limit. These are the terms $\mathcal{L}^n$ (or $\mathcal{K}^{n-1}$), with $n>d$. For $D=4$ there are only two such terms. Explicitly, the $\mathcal{L}$-forms read:
\bea\mathcal{L}^3 (d=2)   &=&\f{1}{2}\pi_\mu\pi_\nu G^{\mu\nu}\,,\label{L3}\\
\mathcal{L}^4(d=3)&=&\f{1}{4}\left(\pi_\mu\pi_\nu\pi^{\mu\nu}-(\partial\pi)^2\Box\pi\right)R-
\f{1}{2}\pi_\mu\pi_\nu\pi_{\rho\sigma}R^{\mu\rho\nu\sigma}\nonumber\\
&&\hspace{-0.3cm}+\f{1}{2}\left(\pi_\mu\pi_\mu\Box\pi+(\partial\pi)^2\pi_{\mu\nu}
-2\pi_\mu\pi_\rho\pi^\rho_{\,\,\nu}\right)R^{\mu\nu}\,.\label{L4}\eea
The first of these extra Galilean terms appeared already in several papers. In fact, in \cite{MuellerHoissen:1989yv,Kobayashi:2004hq,Amendola:2005cr} this term was obtained in a special case of our general procedure, through a toroidal compactification or an equivalent dimensional reduction of 5D Gauss-Bonnet gravity. Note that this specific compactification also generates the original Galileon (\ref{K1}) and of course the non-minimal Gauss-Bonnet coupling (\ref{piGB}). More recently, the term (\ref{L3}) was constructed directly, by demanding second-order equations \cite{Sushkov:2009hk}. To our knowledge, the second extra Galilean term (\ref{L4}) is new to the literature.  
\section{Conclusions}
The original motivation for this paper was to connect the general covariant Galileons of \cite{DDE} to the Lovelock invariants. In our case, the connection arises through standard Kaluza-Klein compactifications of the different pure gravity Lovelock terms. With the dilaton field $\pi$ playing the role of the Galileon, such compactifications reproduce exactly all the covariant Galileon actions of DDE. In addition, our Lovelock construction produce another equally elegant form, for which the equations are identical to those generated by the original Galileon actions. Taken together, Galileons have now been produced for general dimension $D$, either through explicit construction by DDE, or through Lovelock compactifications, and this in two forms (our $\mathcal{L}^d$ and $\mathcal{K}^{d-1}$). Moreover for $D=4$, they were obtained through a similar, but not identical, explicit construction in \cite{DEV} and through a brane setup in \cite{dRT}. The fact that all different constructions lead to identical expressions, clearly speaks in favor of their purported uniqueness.

Furthermore, our Lovelock setup does not only reproduce the covariant Galileons, it also generates a new class of scalar-tensor couplings that lead to second-order equations. And in the weak field limit, which in this case is of nonzero order in the metric perturbations, the equations are purely second-order. In that sense these terms are Galilean,  and we can interpret them as interpolations between the scalar field Galileons and the pure gravity Lovelock invariants. Interestingly, for $D=4$ there exist only two of these scalar-tensor couplings (\ref{L3},\ref{L4}). They add to the well known and well explored Ricciscalar (\ref{piR}) and Gauss-Bonnet (\ref{piGB}) couplings. The phenomenology of the term (\ref{L3}) is already explored to some extent \cite{Amendola:2005cr,Sushkov:2009hk,Amendola:2007ni,Saridakis:2010mf,Germani:2010gm}. It should be worthwhile to explore the new term (\ref{L4}) as well. 

Since their introduction, several variations on the Galileons have appeared. Multi-field Galileons for instance, generalize the Galilean character to an arbitrary number of scalar fields, possibly connected with an extra symmetry \cite{multi-field}. We expect our construction to carry through directly to these new models. Indeed, allowing more scalar degrees of freedom in the parameterization of the extra-dimensional metric $g_{ab}$ in (\ref{metric}), should produce the different (covariant) multi-field Galileons. Another obvious extension of our construction would be to consider the vectors $A_{\mu a}=g_{\mu a}$, in the off-diagonal components of the metric. This should then produce Galilean actions that include several vector fields in addition to the scalars. Actions of this type were constructed in \cite{DDE2} as a special case of arbitrary $p$-form Galileons. However, we do not see an immediate generalization of our method to the general case $p>2$.

Finally, let us stress that in this paper we have merely used Lovelock compactifications as a tool to generate individually interesting terms. For generic compactifications of general combinations of different Lovelock invariants, the coefficients for all different terms in the action will be order one. This makes standard Lovelock compactifications rather unnatural as a physical mechanism for Galilean IR modified gravity, which only requires the original Galilean terms (\ref{K0}-\ref{K3}), together with the Ricciscalar (\ref{R}) and Ricciscalar coupling (\ref{piR}). To recover ordinary gravity at short distances, the coefficients of the other terms like (\ref{piGB}-\ref{L4}) would have to be heavily suppressed. This is possible in principle, but only at the cost of some fine-tuning of both the Lovelock combinations and the extra-dimensional curvature. On the other hand, there is no immediate objection against these extra scalar-tensor couplings in the context of Galilean inflation.
\appendix
\numberwithin{equation}{section}
\section{Appendix}
\subsection{Decomposition of $L_{(d)}^{D+N}$}\label{decomp}
We straigtforwardly calculate the Riemann curvature $R$ corresponding to the $D+N$ -dimensional metric ansatz (\ref{metric}). Using $R$ also for the "ordinary" $D$-dimensional curvature and denoting $R^{(N)}$ for the one from the $N$-dimensional $\gamma_{ij}$, we have:
\begin{eqnarray}
R^{\beta_1\beta_2}_{\alpha_1\alpha_2}&=&
g^{\beta_1\gamma_1}g^{\beta_2\gamma_2}R_{\gamma_1\gamma_2\alpha_1\alpha_2}\sim R\nonumber\,,\\
R^{d\delta}_{c\gamma}=-R^{\delta d}_{c\gamma}=R^{\delta d}_{\gamma c}=-R^{d \delta}_{\gamma c}&=&-\frac{1}{2}\delta_c^d(\frac{1}{2}\pi_\gamma \pi^\delta+\pi_\gamma^\delta )\sim S\label{Theta}\,,\\
R^{b_1b_2}_{a_1a_2}&=&e^{-\pi}R^{(N)b_1b_2}_{a_1a_2}-\frac{1}{4}\delta^{b_1b_2}_{[a_1a_2]}(\partial\pi)^2\sim T\,,\nonumber
\end{eqnarray}
with the other components zero. 

Our first aim will be to decompose the $d$-th Lovelock-invariant as a formal polynomial in the above $R$, $S$ and $T$:
\be L_{(d)}^{D+N}\sim\sum R^{p}S^{q+1}T^l\,.\ee
Notice that this is, keeping $d$ fixed, only a double sum as $d=l+p+q+1$. The "$+1$" owing to the conventions of \cite{DDE}. For later convenience we also introduce $n=2p+q+1$, the total number of (upper) greek indices, and $m=2l+q+1$, the total number of (upper)latin indices in the particular term of the Lovelock $\delta$-tensor, cfr. (\ref{LL}).
It takes some combinatorial effort to see how these terms can be rearranged:
\begin{eqnarray}
L_{(d)}^{N+D}&=&\delta^{A_1\cdots A_{2d}}_{[B_1\cdots B_{2d}]}R_{A_1A_2}^{B_1B_2}\ldots\nonumber\\
&=&\sum_{\sigma\in S_{2d}}sgn(\sigma)\delta^{A_1}_{B_{\sigma(1)}}\cdots\delta^{A_{2d}}_{B_{\sigma(2d)}}R_{A_1A_2}^{B_1B_2}\ldots\nonumber\\
&=&\sum_{l,p} \frac{d!2^{q+1}}{p!(q+1)!l!}\sum_{\sigma\in S_{2d}}sgn(\sigma) \delta^{a_1\cdots c_1\cdots \gamma_1\cdots\alpha_{2p}}_{B_{\sigma(1)}\cdots \cdots\cdots B_{\sigma(2d)}}T^{l}S^{q+1}R^{p}\,,
\end{eqnarray}
with $\sigma$ denoting a permutation of signature $sgn(\sigma)$ of the permutation group $S_{2d}$ and suitable boundaries for the $l,p$ summations implied.  Notice the symmetry factor $2^{q+1}$ on the last line, that compensates for fixing the positions of $c_i$ and $\gamma_i$ in the $S$-tensors.
As all terms containing a Kronecker-delta $\delta^{greek}_{latin}$ vanish, only a sum over $\sigma\in S_n\times S_m$ remains, and so the Lovelock $\delta$-tensor factors out in one for latin and one for greek indices:
\be L_{(d)}^{N+D}=\sum_{l,p} \frac{d!4^{q+1}}{p!(q+1)!l!}\delta^{a_1\ldots  c_1\ldots c_{q+1}}_{[b_1\ldots d_1\ldots d_{q+1}]}\delta^{\gamma_1\ldots\alpha_1\ldots\alpha_{2p}}_{[\delta_1\ldots\beta_1\ldots\beta_{2p}]}
\prod^{l}_{i=1}T_{a_{2i\sm1}a_{2i}}^{b_{2i\sm1}b_{2i}}
\prod_{j=1}^{q+1}S^{d_j\delta_j}_{c_j\gamma_j}\prod_{k=1}^{p}R_{\alpha_{2k\sm1}\alpha_{2k}}^{\beta_{2k\sm1}\beta_{2k}}
\,.\ee
Again, there is an extra symmetry factor $2^{q+1}$, that now compensates for fixing the positions of $d_i$ and $\delta_i$ in the $S$-tensors.
Further simplification follows from the identity:
\be
\delta^{a_1\cdots a_{2l}c_1\cdots c_{m-2l}}_{[b_1\cdots b_{2l}c_1\cdots c_{m-2l}]}=\frac{1}{(N-m)!}\delta^{a_1\cdots c_{m-2l}d_1\cdots d_{N-m}}_{[b_1\cdots c_{m-2l}d_1\cdots d_{N-m}]}
=\frac{(N-2l)!}{(N-m)!}\delta^{a_1\cdots a_{2l}}_{[b_1\cdots b_{2l}]}\,,\label{iddelta}
\ee
and the fact that $S\sim\delta$ for the latin indices, so:
\be
L_{(d)}^{N+D}\sim\sum_{l,p} \frac{d!(N-2l)!}{p!(q+1)!l!(N-m)!}4^{q+1}
\delta^{a_1\ldots a_{2l}}_{[b_1\ldots b_{2l}]}\delta^{\gamma_1\ldots\gamma_{q+1}\alpha_1\ldots\alpha_{2p}}_{[\delta_1\ldots\delta_{q+1}\beta_1\ldots\beta_{2p}]}
T^{l}S^{q+1}R^{p}\,.\ee
In a similar way we can reduce the sum of the remaining latin indices, identifying the  
$N$-dimensional order $r$ Lovelockterms $L^N_{(r)}$ hidden in $T^l$:
\be\delta^{a_1\ldots a_{2l}}_{[b_1\ldots b_{2l}]}T^l=\sum^l_{r=0}\frac{l!(N-2r)!}{(l-r)!r!(N-2l)!}e^{-r\pi}L^N_{(r)}(-\frac{1}{2}\pi_{\mu}\pi^{\mu})^{l-r}\,.\ee
And thus finally, taking into account that antisymmetry in the indices will allow only the first two terms in the binomial expansion of $S^{q+1}$:
\begin{align}
L_{(d)}^{N+D}&=\nonumber\\
&{}\sum_{l,p,r}\mathcal{C}e^{\sm r\pi}L^N_{(r)}(\pi_{\mu}\pi^{\mu})^{l\sm r}\delta^{\gamma_1\ldots\alpha_{2p}}_{[\delta_1\ldots\beta_{2p}]}\left(\pi_{\gamma_1}^{\delta_1}+\frac{q+1}{2}\pi_{\gamma_1}\pi^{\delta_1} \right)\prod_{i=2}^{q+1}\pi_{\gamma_i}^{\delta_i}\prod_{j=1}^pR^{\beta_{2j\sm 1}\beta_{2j}}_{\alpha_{2j\sm 1}\alpha_{2j}}\nonumber\\
={}&\sum_r e^{-r\pi}L^N_{(r)}\sum_{n}\mathcal{\tilde C}^n\sum_p\left(\mathcal{D}_p\mathcal{K}^{n}_{p}+\mathcal{C}_p\mathcal{L}^{n}_{p}\right)\label{FF}\\
\equiv{}&\sum_r e^{-r\pi}L^N_{(r)}\sum_{n}\mathcal{\tilde C}^n\left(\mathcal{K}^{n}+\mathcal{L}^{n}\right)\,,\nonumber
\end{align}
with
\begin{eqnarray}
\mathcal{C}&=& \frac{d!}{p!(q+1)!l!}(-2)^{q+1}\frac{(N-2l)!}{(N-m)!}\frac{l!}{(l-r)!r!}\frac{(N-2r)!}{(N-2l)!}(-2)^{r-l}\nonumber\\
&=&\frac{2}{q+1}\left(\frac{(-2)^{2n+r-d}}{2}\frac{d!}{r!}\frac{(N-2r)!}{(N+n-2d)!}\right)\left(\frac{(-2)^{-3p}}{p!(d-n+p-r)!(n-2p-1)!}\right)\nonumber\\
&=&\frac{2}{n-2p}\mathcal{\tilde C}^n\mathcal{C}^n_p=\mathcal{\tilde C}^n\mathcal{D}^n_p\,\label{coeff}.
\end{eqnarray}

On the second line of (\ref{FF}), we changed the summation index $l\rightarrow n$. One can easily verify that the final summation bounds are $0 \leq r \leq d$ for the $r$-summation, while the $n,p$- summation simply runs over all nonzero terms, given the coefficients (\ref{coeff}).

Finally, naming all $2n$ indices in the remaining Lovelock-$\delta$ $\mu_i$, such that the $\alpha$'s and $\gamma$'s get odd $i$ and $\beta$'s and $\delta$'s even $i$, we get exactly the form (\ref{A2n}) introduced in section \ref{sec2}, and can adopt the notation of \cite{DDE}:
\begin{eqnarray}
\mathcal{K}^{n}_{p}&=&\mathcal{A}_{(2n)}\mathcal{S}(q+1)(\pi_{\nu}\pi^{\ \nu})^{l-p-r}\mathcal{R}(p)\,,\\
\mathcal{L}^{n}_{p}&=&\mathcal{A}_{(2n)}\pi_{1}\pi_{2}\mathcal{S}(q)(\pi_{\nu}\pi^{\ \nu})^{l-p-r}\mathcal{R}(p)\,.
\end{eqnarray}

\subsection{$\mathcal{K}/\mathcal{L}$ ambiguity}\label{divergence}
To verify relation (\ref{KL}), we first of all introduce $c=d-r-n$ for notational simplicity. One can then rewrite the linear combination $\frac{n+2c}{2}\mathcal{K}^n-2\mathcal{L}^{n+1}$ by using the explicit form of the coefficients (\ref{coeff}):
\begin{multline}
\mathcal{A}_{(2n)}\sum_{p}\left(\mathcal{C}_{p}^{n}+2\mathcal{C}_{p}^{n+1}\right)\mathcal{S}(q+1)(\partial\pi)^{2c}\mathcal{R}(p)\label{previous}\\
-2\mathcal{A}_{(2n+2)}\sum_p\mathcal{C}_{p}^{n+1}\left(\pi_{2n+1}\pi_{2n+2}\pi_{12}R_{3546}\right)\mathcal{S}(q)(\partial\pi)^{2c}\mathcal{R}(p-1)
\,.\end{multline}
We now factor out all the terms $g^{\mu_{2n+1}\mu_{2i}}$ in $\mathcal{A}_{(2n+2)}$. The highly symmetrical form of the contracted expression reduces these $n+1$ contributions to 3 different forms, in which the index of $\pi_{2n+1}$, is contracted with an (even) index of either $\pi_{2n+2}$, $\pi_{12}$ or $R_{3546}$, respectively. The remaining cofactor is always, after rearranging indices, $\pm\mathcal{A}_{(2n)}$:
\begin{multline}
\mathcal{A}_{(2n+2)}\left(\pi_{2n+1}\pi_{2n+2}\pi_{12}R_{3546}\right)\mathcal{S}(q)\mathcal{R}(p-1)=\\
\Bigl(\mathcal{A}_{(2n)}\pi_\kappa\pi^\kappa\pi_{12}R_{3546}-(q+1)\mathcal{A}_{(2n)}\pi_2\pi^\kappa\pi_{1\kappa}R_{3546}\\+2p\mathcal{A}_{(2n)}\pi_2\pi^\kappa\pi_{14}R_{35\kappa6}\Bigr)\mathcal{S}(q)\mathcal{R}(p-1)\,.
\end{multline}
Putting all this in (\ref{previous}), we find the first term here canceling the second term on the first line of (\ref{previous}),  while we will alter the dummy  $p\mapsto p+1$ in the last term. This yields for $\frac{n+2c}{2}\mathcal{K}^n-2\mathcal{L}^{n+1}$:
\begin{multline}
\mathcal{A}_{(2n)}\sum_{p}\Bigl\lbrack\mathcal{C}_{p}^{n}\mathcal{S}(q+1)(\partial\pi)^{2c}\mathcal{R}(p)+2\mathcal{C}_{p}^{n+1}(q+1)\pi_2\pi^\kappa\pi_{1\kappa}\mathcal{S}(q)\mathcal{R}(p)(\partial\pi)^{2c-2}
\\
-2\mathcal{C}_{p+1}^{n+1}2(p+1)\pi_2\pi^\kappa\pi_{14}R_{35\kappa6}
\mathcal{S}(q-2)\mathcal{R}(p)(\partial\pi)^{2c}\Bigr\rbrack
\,,\end{multline}
which, using again the explicit form of the coefficients (\ref{coeff}), becomes: 
\begin{multline}
\sum_p\mathcal{C}_{p}^{n}\left(\pi_{12}\pi_{34}(\partial\pi)^{2}+2(c+p)\pi_2\pi^\kappa\pi_{1\kappa}\pi_{34}+\frac{q}{2}\pi_2\pi^\kappa (\partial\pi)^{2}R_{31\kappa4}\right)\\
\mathcal{S}(q-1)\mathcal{R}(p)(\partial\pi)^{2c-2}\,.
\end{multline}
Finally, it is easily verified, using the anti-symmetry of $\mathcal{A}_{(2n)}$ in combination with the Bianchi identity,  that this last expression is nothing but $\nabla_1\left(\pi_2\mathcal{A}_{(2n)}\sum_p\mathcal{C}_p^n\mathcal{S}(q)\mathcal{R}(p)(\partial\pi)^{2c}\right)$, proving the relation:
\be\frac{n+2c}{2}\mathcal{K}^n-2\mathcal{L}^{n+1}=\nabla_1\left(\pi_2\mathcal{A}_{(2n)}\sum_p\mathcal{C}_p^n\mathcal{S}(q)\mathcal{R}(p)(\partial\pi)^{2c}\right)\,.\ee

\bibliographystyle{plain}

\end{document}